# Affective State Recognition through EEG Signals Feature Level Fusion and Ensemble Classifier

**Md. Mahbubur Rahman** *[1], **Akash Poddar** [2], **Md. Golam Rabiul Alam**[3], **and Samrat Kumar Dey**[4]

[1,2]Department of Computer Science and Engineering (CSE), Military Institute of Science and Technology (MIST), Mirpur Cantonment, Dhaka-1216, Bangladesh

[3]Department of Computer Science and Engineering (CSE), BRAC University, Dhaka, Bangladesh

[4]Department of Computer Science and Engineering (CSE), Dhaka International University (DIU), Dhaka-1205, Bangladesh

* **Corresponding Author**: Md. Mahbubur Rahman
Present address: Department of Computer Science and Engineering (CSE),
Military Institute of Science and Technology (MIST), Mirpur Cantonment, Dhaka-1216, Bangladesh
Email: mahbubcse@yahoo.com

**E-mail**: mahbubcse@yahoo.com[1]; akash.poddar.0799@gmail.com[2]; rabiul.alam@bracu.ac.bd[3]; and sopnil.samrat@gmail.com[4]

## Abstract

Human affects are complex paradox and an active research domain in affective computing. Affects are traditionally determined through a self-report based psychometric questionnaire or through facial expression recognition. However, few state-of-the-arts pieces of research have shown the possibilities of recognizing human affects from psychophysiological and neurological signals. In this article, electroencephalogram (EEG) signals are used to recognize human affects. The electroencephalogram (EEG) of 100 participants are collected where they are given to watch one-minute video stimuli to induce different affective states. The videos with emotional tags have a variety range of affects including happy, sad, disgust, and peaceful. The experimental stimuli are collected and analyzed intensively. The interrelationship between the EEG signal frequencies and the ratings given by the participants are taken into consideration for classifying affective states. Advanced feature extraction techniques are applied along with the statistical features to prepare a fused feature vector of affective state recognition. Factor analysis methods are also applied to select discriminative features. Finally, several popular supervised machine learning classifier is applied to recognize different affective states from the discriminative feature vector. Based on the experiment, the designed random forest classifier produces 89.06% accuracy in classifying four basic affective states.

## 1.0 Introduction

Affective state or human emotion is a mental state linked to the nervous system. Therefore, analyzing signals from human nervous systems would be the source of affective state recognition. The electroencephalogram (EEG) produces the activity signals of the brain which is the central part of the human nervous system. All these signals are called brain waves and can be distinguished by the features they carry. Thus the brain waves carry characteristics of the emotion and the emotional state of the brain or human can be identified by analyzing the brain wave. Identification of emotion using brain waves can be utilized for different purposes. In medical science, this can be used to mine moods for detecting bipolar disorders. In the industrial sector, the level of job satisfaction of the workers can be measured by analyzing their affective state. There had been much research in the field of emotion detection using facial expression recognition. Yet, true emotion may not be extracted from artificial or fake facial expressions. Some of the earlier researches **[1] [2]** utilized 32 channel EEG and other peripheral sensors to detect emotion. However, those lab experiments with too many sensors are not suitable for real-time usage. Conversely, in the proposed emotion detection system we utilized two channels EEG for collecting brain waves. The system can be used in real-time environment and very easy to operate. Human emotions create physiological signals which are generated from the brain. These are incorporated with thoughts, feelings, behavioural responses, and a degree of pleasure or displeasure **[3]**. The emotions of ours can impel us to take action and dominate the decisions in lives. Emotion can be referred as joy, anger, disgust, sadness, fear and surprise of human feelings. In other words, Emotion recognition is a method used to use hi-tech image processing tools to read the emotions on a human face. Human-computer interaction (HCI) is computer technology, focused on the interfaces between humans and computers. There are different approaches of extensive scales of emotion, such like: Plutichik's emotion wheel **[4]**, valence-arousal scale by Russell et al., **[5]**. According to this process, each emotional state can be plotted into a two-dimensional plane where horizontal and vertical axes represent arousal and valence respectively. This system requires distinct features to be plotted in the plane. There are many brain signals emitted from the brain and only very few are utilized in the above-stated system. In our proposed system, total of seven types of signals are extracted from the device. However, not all the signals carry distinct characteristics those can be analysed and the better system can be brought into the light. There had been many challenges that research team faced in different stages of this research. One of the major challenge is a noise-free lab environment to get a strong dataset. For the data collection, a noise-free lab environment was not available. There are only two channels in the device to read the brain waves. This fact is a limitation for this experiment as other devices got a number of brain wave reading channels. However, it is otherwise a requirement of the research as this fact makes this methodology more acceptable in the industrial utilization of the workers. The objectives of this research are to recognize real-time emotions from two-channel EEG brain waves by extracting advanced and discriminative EEG features for differentiating affective states. Also, to perform feature level fusion of statistical features and advanced EEG features for the vast discriminative feature vector generation by applying factor analysis methods for selecting discriminative features, and finally, to design a multiple supervised learning classifier to classify affective states efficiently.

## 2.0 Relevant Work

Currently, researches are conducted on emotion recognition or classification based on physiological signals or speech or face expression image processing. Koelstra et al. **[2]** has formulated a vast DEAP dataset based on EEG and peripheral physiological signals of 32 participants where each of them watched several videos. Authors have worked with classification of arousal, valence, and like/dislike ratings of the participants. Mavani et al. **[6]** in his research has worked on CFEE and RaFD datasets by improving the existing CNN model and could reach accuracy of 65.39%. Another work based on physiological signals like EDA, PPG, zEMG are fused to differentiate five major emotion classes using Fina Gaussian Support

Vector Machine in [7]. Apart from that to extract features, authors have applied Deep Belief Network (DBN). They were able to reach at 89.53% accuracy. Tripathi et al. [8] have also worked on DEAP dataset classifying emotion using EEG signals using both Deep Neural Network and Convolutional Neural Network (CNN) proving effectiveness over existing work on this arena. Another research on signal based emotion recognition is conducted by Alam et al. [9] by using EMG, EDA, ECG sensors are analyzed using deep CNN with accuracy of 87.5%. Meanwhile, Mohammadi et al. [10] used SVM and K-nearest neighbor classifiers to detect emotion using 10-channel EEG signal. Yet, researchers have worked with various sensors but their quality and emotion detection capabilities had some concern in their research work. To check the grade and potentiality of lab based and wearable sensors Ragot et al. [11] compared their accuracy. However this experiment proved reliability of both types of sensors. In other experiment, Zhuang et al. [12] used EMD and EEG signals for feature extraction and emotion recognition decomposing into empirical mode decomposition (EMD). Their multidimensional information is used as features. The authors have checked their accuracy of classification comparing with several classical techniques; including fractal dimension (FD), sample entropy, differential entropy, and discrete wavelet transform (DWT). However, many research works has been carried out on EEG signals but its stability over time remains unexplored. The frequently used popular feature extraction, feature selection, feature smoothing and pattern classification methods are analyzed and evaluated in [13] using public dataset DEAP and their own developed dataset SEED. The emotion recognition model shows that the neural patterns are relatively stable within and between sessions. Xia et al. [13] through their research has enforced activation and valence information for acoustic emotion recognition applying multi-task learning based on DBN. Besides, authors have enforced activation and valence in two different ways: category level based classification and continuous level based regression. The fusion of the loss functions from both tasks is used as the objective function in the multi-task learning framework. After iterative optimization, the values from the last hidden layer in the DBN are used as new features and made input into a support vector machine (SVM) classifier for emotion recognition. Though conventional methods of using EEG signals for emotion recognition skips the use of spatial characteristics of EEG signals which contain various salient features, Chao et al. [14] have used frequency domain, spatial characteristics, and frequency band characteristics of the multichannel EEG signals to build up multiband feature matrix (MFM). Based on input MFM, a capsule network (CapsNet) classifies emotion states. Ullah et al. [15] has proposed an ensemble learning algorithm for automatically computing the most discriminate subset of EEG channels for internal emotion recognition. This method describes an EEG channel using kernel-based representations computed from the training EEG recordings. This algorithm reduces the amount of data along with improving computational efficiency and classification accuracy at the same time. A complete different arena of emotion detection is represented on [16]. It works with emotion detection from surveillance camera video. The body shape and gesture from video identifies the emotion which is done on basis of publicly available dataset and modeled by Support Vector Machine (SVM) with accuracy of 93.39%. Authors in **[17]** have performed continuous emotion prediction on three dimensions: Arousal, Valence, and Likability based on audiovisual signals. Authors also have measured the contrast between effectiveness of non-temporal model SVR and temporal model LSTM-RN. In recent time, Li et al. **[17]** have developed a series of EEG Multidimensional Feature Image (EEG MFI) sequence from special characteristics, frequency domain and temporal characteristics mapping them into two-dimensional image. Besides, they have also constructed hybrid deep neural network along with CNN, LSTM, RNN to work out with EEG MFI sequence to classify human emotion and finally got accuracy over 75.21%. Based on the discussion it is evident that, Human facial expression is not the superlative way to determine human emotion. This is because a person might be mentally sad but showing pseudo happy emotional expression through his face. Thus working with facial emotion recognition might decrease the accuracy rate of the work. The best way to deal with this problem is physiological signals. Brain impulse are used here which is EEG signal and it generates different types of signals for different types of emotions. As a result, this process of emotion classification is more trustworthy and provides better accuracy than other processes. At the same time, our research team have tried to find out the accuracy of classifying emotion correctly only using two channels. Here the four (4)-emotion states including funny, sad, disgust and peaceful are considered and a customized dataset is collected based on the response of 100 persons. Based on the dataset, statistical feature extraction is conducted. At the final stage, depending on various machine-learning algorithms the accuracy result of testing is analyzed.

## 3.0 Methodology

The system model of the proposed affective state recognition framework is presented in **Figure 1**. This section discussed the detail methodology of the proposed system.

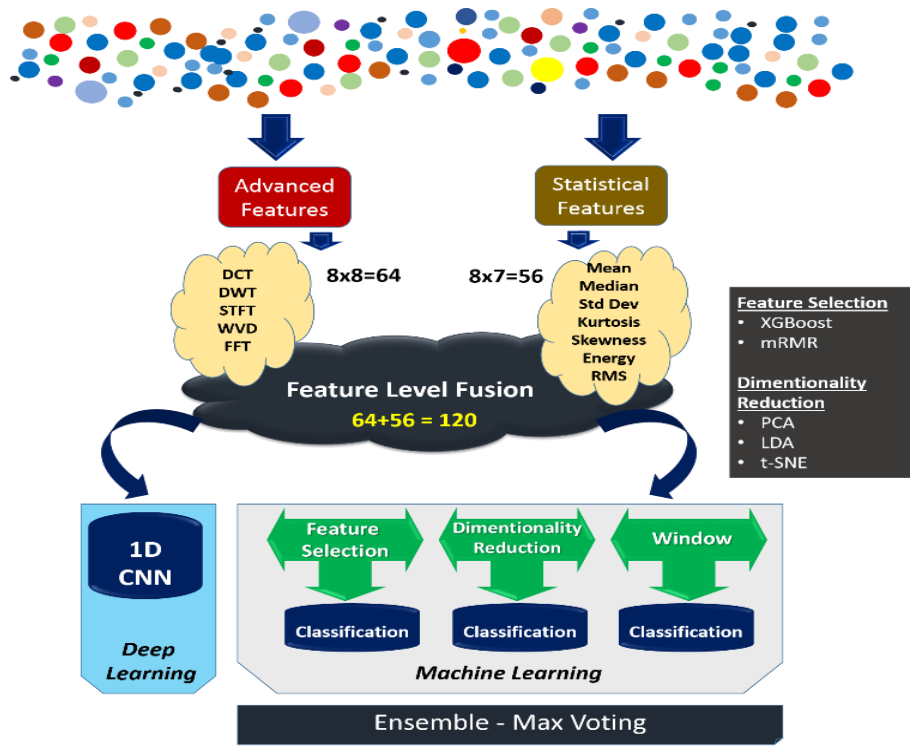

**Figure 1.** The system model of the proposed affective state recognition framework

## 3.1 Data Collection

We have prepared our own dataset for the analysis of human affective states. The electroencephalogram (EEG) signals of 100 participants were recorded as each watched one-minute long videos of four different subjects. The objective was to record the values for four different emotion state. The stimuli used in the experiment were selected in several steps. First, we have selected 15 initial stimuli manually. Then, a one-minute highlight part was determined for each stimulus. Finally, through the web-review process we have examined the content type. The videos were downloaded from famous YouTube channels. Based on the result of the google engine for various tags, we have selected the most rated ones. The valence-arousal space to work on emotion classification can be split into four (4) regions: low arousal/low valence (LALV), low arousal/high valence (LAHV), high arousal/low valence (HALV) and high arousal/high valence (HAHV) as illustrated in **Figure 2**. Based on linear regression method, arousal and valence in each movies was determined. Loudness and energy of the audio signals, motion component, visual excitement and shot duration were considered for arousal calculation. The regressor were trained using the dataset for better valence and arousal estimation.

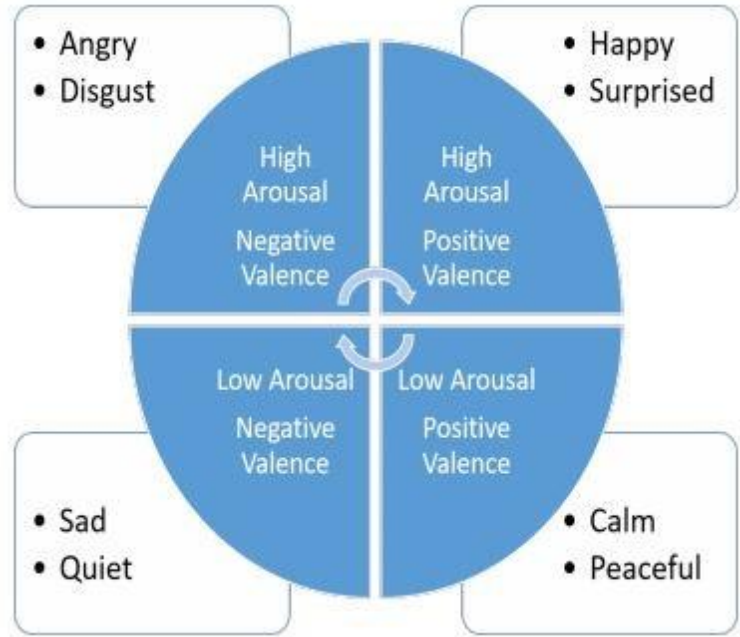

**Figure 2:** Arousal Valence Scale of Emotion

This experiment have utilized the concept of Relevance Vector Machine (RVM) from RVM toolbox, which is capable of refusing the vague features throughout the training process. We have employed the Mindwave Mobile 2 android application for the experiment. The electric signal emitted by neurons fired in the brain has different patterns and frequencies. This signals are measured by placing sensors in the scalp. This Mindwave NeuroSky Gear measures the Analog electrical signals, commonly referred to as brainwaves, and processes them into digital signals. The frequencies that Mindwave application generally capable of reading are shown in **Table 1**.

**Table 1:** Mind wave Channel Types

| Brainwave Type | Frequency range | Mental State and conditions |
|---|---|---|
| Delta | 0.1 Hz to 3 Hz | Deep, dreamless sleep, non REM-sleep, unconscious |
| Theta | 4 Hz to 7 Hz | Intuitive, creative, recall, fantasy, imaginary, dream |
| Alpha | 8 Hz to 12 Hz | Relaxed (but not drowsy) tranquil, conscious |
| Low Beta | 12 Hz to 15 Hz | Formerly SMR, relaxed yet focused, integrated |
| Midrange Beta | 16 Hz to 20 Hz | Thinking, aware of self & surroundings |
| High Beta | 21 Hz to 30 Hz | Alertness, agitation |

This device has a sensor that touches the forehead, has contact points in the ear clip and on board chip that processes all the data. It has proprietary algorithms for characterizing mental states. During calculation of the data, it amplifies the raw brainwave signal and removes the ambient noise and muscle movement. Mindwave Mobile 2 android application is properly setup and used for the collection of the signals/data from EEG sensors. Since our study focuses on classifying four different emotions, therefore finally selected four stimuli videos were shown individually to the viewers. Each video continued for 60 seconds. We have collected the emotions of 100 different individuals. Therefore, the total number of data is 60 x 4 x 100 = 24000. The notations we have followed for the emotions are shown in **Table 1**. In the raw data there had been few columns which does not have any significance with our research interest. Therefore, primarily we have considered 12 columns (features) and the dimension of our dataset has become 24000 x 12 = 288000 values.

## 3.2 Data Pre-processing

After the completion of data collection, the next step that have been followed is pre-processing the data. Data rescaling is an important part of data preparation before applying machine-learning algorithms. It enhance the result and reduces unnecessary data. Therefore, the steps followed by the research group is selection of significant features, reduction of overfitting data, and normalization. In this step of selection of significant features, we have basically reduced the feature size for our experiment. The features which shows significance with the research objectives, we have kept those for detail investigation. Out of initial 12 features, we have observed the relevance of eight (8) features (delta, theta, alphaLow, alphaHigh, betaLow, betaHigh, gammaLow, gammaMid). As such, data size dimension has reduced to (row x column) = 24000 x 8 = 192000 values. In the step of reduction of over fitting data, we have removed the unnecessary data that were identified as irrelevant bases on their values. Moreover, our data may contain attributes with a mixture of scales for various quantities. Since, many machine-learning methods are more effective if the data attributes have the same scale therefore, we have also utilized data normalization process. Here, due to normalization, all the values are set in between 0 and 1, and the outliers are also removed as well. All our features are more consistent with each other, which will allow us to evaluate the output of our future models more efficiently.

## 3.3 Feature Extraction

After the following statistical feature applied to the dataset, total number of features stands 8 (raw features) x 7 (statistical features) = 56 (7 Statistical feature applied to each raw feature) and for one type of emotion one person contains 56 features. Similarly, for one emotion 100 persons has 100 x 56 emotions. It indicates that, for four (4) different emotions the dimension has become (row x column) 400 x 56 = 22400 values. The arithmetic mean is the average of the values located within a time window. The median is the middle value when a data set is ordered from least to greatest. Following **Equation (1)** represents the distortion or asymmetry in a symmetrical bell curve, or normal distribution, in a set of data.

$$Skewness = E[\frac{(x-\mu)^3}{\sigma}] \dots\dots\dots\dots\dots\dots\dots\dots\dots\dots\dots\dots\dots\dots\dots\dots\dots\dots\dots (1)$$

There is a skew of zero in a normal distribution, while, for example, a lognormal distribution would show some degree of right skew.

$$K(s) = E(s^4) - 3E(s^2)^2 \dots \dots \dots \dots \dots \dots \dots \dots \dots \dots \dots \dots \dots \dots \dots \dots \dots \dots \dots \dots \dots \dots \dots \dots \dots (2)$$

Coefficients of EEG signal do not follow the normal distribution, and have a heavy tail characteristic which is justified by the value of kurtosis parameters. Positive kurtosis indicates a relatively peaked distribution whereas negative kurtosis indicates a relatively flat distribution. **Equation (2)** shows the mathematical formula of Kurtosis. Standard deviation is the measure of spread. It measures spread around the mean. Because of its close links with the mean, standard deviation can be greatly affected if the mean gives a poor measure of central tendency. Entropy is a statistical model of the signal which can also provide physical information. It measures the signal complexity and quantify regularity and order in the signal. It is observed that low entropy value of EEG signals represents less number of dominating process and the EEG signals with high entropy represent large number of dominating processes. After applying the above features the present dataset is saved for further cross references. For the better outcome of calculation and performance evaluation, further study has been conducted for advanced features measurement. Thereby, total of 5 and other 3 (mixture of five features) total 8 features are taken. And each feature has 8 different sub features. So, after considering the advanced features over our raw dataset the dimension has stand to (row x column) 400 (for 4 emotions) x 64 = 25,600 values. Short Time Fourier Transform (STFT) extracts several frames of the original signal to be analysed with a window frame that shifts with time **(Equation 3)**.

$$S_\omega(t) = \frac{1}{\sqrt{2\pi}} \int e^{j\omega' t} S(\omega') H(\omega - \omega') d\omega' \dots \dots \dots \dots \dots \dots \dots \dots \dots \dots \dots \dots \dots \dots \dots \dots \dots (3)$$

For analysis and modelling of slowly changing signals, Fourier analysis have been widely used. To study frequency characteristics at moments STFT is examined. Conversely, at a given frequency it is also possible to examine the time characteristics. Discrete Wavelet Transform (DWT) is the process of transforming time signal to a discrete wavelet representation.

$$y[n] = (x * g)[n] = \sum_{k=-\infty}^{\infty} x[k] g[n-k] \dots \dots \dots \dots \dots \dots \dots \dots \dots \dots \dots \dots \dots \dots \dots \dots \dots (4)$$

For processing Biological signals such as EEG in discrete wavelet transform digital filtering is used. DWT algorithm gives octave-scale frequency as well as spatial timing of the given signal **(Equation 4)**. For this reason it is always used to address and resolve many sophisticated and complicated issues. In this process, Low Pass Filter (LFP) and High Pass Filter (HPF) are extracted. Scaling function can be acquired from the LPF while the wavelet function can be acquired from the HPF. The filter bank level varies with the availability of the bandwidth. Discrete Cosine Transform (DCT) is used primarily for signal extraction features. It is a way for transforming a frequency into different frequencies for summation of cosine functions. In this method the relevant coefficients of a signal is transformed from a whole signal. It performs energy compaction and de-correlate the data for the image. After decorrelation of every data, the coefficient can be cipher individualistic. The transformed signal is categorized in low, mid and high frequency each contains different details and information about the signal. Fast Fourier Transform (FFT) is a kind of transforming from the basic Fourier that is much quicker. This is used to turn a signal from the time domain into a frequency domain. The inverse FFT does the opposite conversion. FFT is mostly used for signal processing as it consumes significantly less time than other feature extraction methods **(Equation 5)**. It is an efficient algorithm of the (FT).

$$X[k] = X_1(k) + e^{\frac{-2\pi jnk}{N}} X_2(k)$$ …………………………………………………. (5)

This study group has added parts of the signal product to get Wigner distributed multiplied at some point at a certain point at a certain time by the signal, since past was equal to future. Therefore, we mentally fold the left part of the signal to the right and check whether if there are a simultaneous overlaps for determining Wigner distribution properties. If there is, then at the time t, those properties will now be present. Wigner-Ville distribution (WVD) compared the information of the signal with its own information at other times and frequencies. For the frequency domain, Wigner distribution in both domains is essentially identical. Another significant point is that the distribution of Wigner is equally weighing the distant times to the close moments. The distribution of Wigner is therefore extremely non local.

### 3.4 Feature Selection

XGBoost is an optimized library for distributed gradient boosting, designed to be extremely powerful, scalable and portable. Under the Gradient Boosting paradigm, it applies machine-learning algorithms. XGBoost offers a parallel tree boost (also known as GBDT, GBM) that easily and reliably addresses several data science issues. The mRMR is an approach to feature selection that aims to pick characteristics with a high class (output) correlation and a low correlation between themselves. The F-statistic can be used for continuous characteristics to calculate the class correlation (relevance) and the Pearson correlation coefficient can be used to calculate the correlation between characteristics (redundancy). After that, by applying a greedy search to maximize the objective function, which is a function of importance and redundancy, features are chosen one by one. MID (Mutual Information Difference criterion) and MIQ (Mutual Information Quotient criterion) representing the difference or the significance and redundancy quotient, respectively, are two widely used forms of the objective function. For temporal data, some pre-processing techniques are needed for the mRMR feature selection approach to flatten temporal data into a single matrix in advance. This may contribute to the loss of potentially valuable information in temporal data (such as temporal order information). **Figure 3** and **Figure 4** highlights the statistical feature selection and advanced feature selection after applying XGBoost feature section approach.

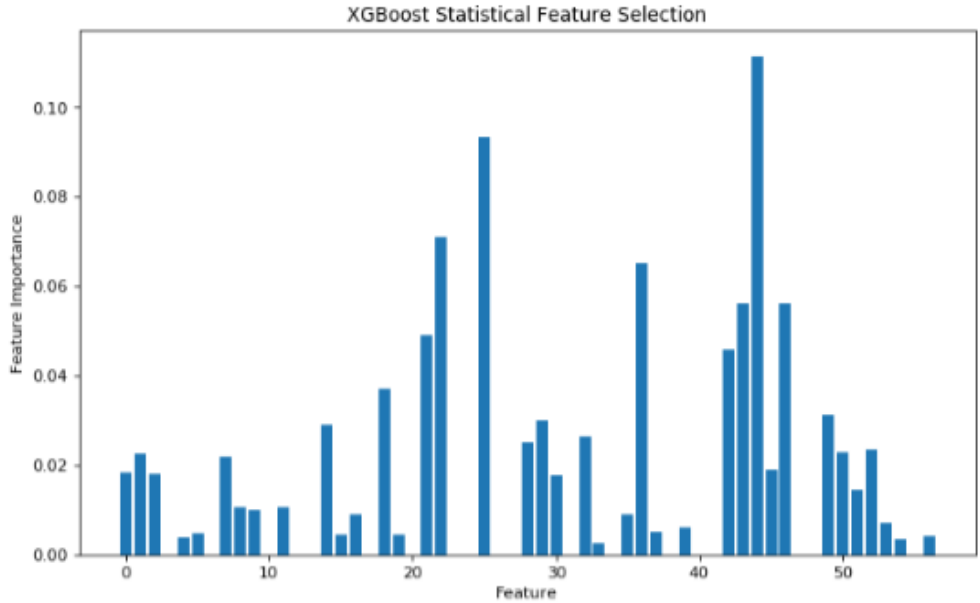

**Figure 3:** Feature importance of statistical feature set through applying XGBoost feature selection method.

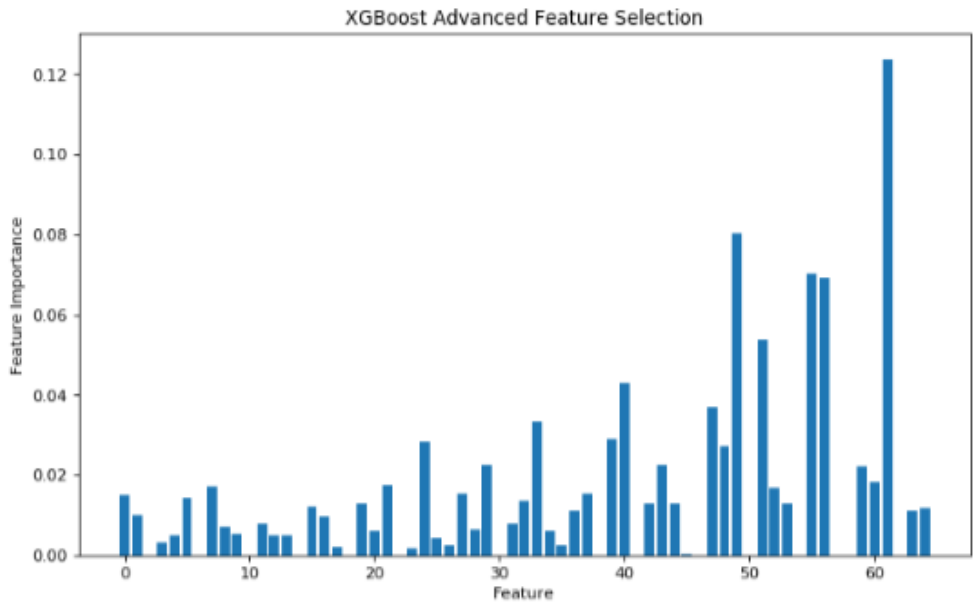

**Figure 4:** Feature importance of advanced EEG feature set through applying XGBoost feature selection method

## 3.6 Dimensionality Reduction

After applying the Principle Component Analysis (PCA), it is found to have 98% of the variance just applying only 20+ statistical features and similarly 98% variance can also be achieved by applying 10+ advanced features. Linear Discriminant Analysis (LDA) is another approach that follows the supervised learning based dimensionality reduction technique. In LDA, it is assumed that the input data follows a Gaussian distribution otherwise; it may possibly lead to poor classification results. We have achieved the **Figure 5** after applying LDA in statistical features. t-distributed Stochastic Neighbour Embedding (t-SNE) is another approach that is basically a non-linear dimensionality reduction technique which is typically used to visualize high dimensional datasets. In t-SNE, a Gaussian distribution is used to model the higher dimensional space, while a Student's t-distribution is used to model the lower-dimensional space. This approach is also carried out in this research to avoid an imbalance in the neighbouring points distance distribution caused by the translation into a lower-dimensional space. SNE starts by converting the high-dimensional Euclidean distances between data points into conditional probabilities that represent similarities. Moreover, it is extremely imperative in capturing both the local and global structure of the highly dimensional data. Instead of looking at directions/axes which maximize information or class separation, t-SNE converts the Euclidean distances between points into conditional probabilities. A student-t distribution is then used on these probabilities which serve as metrics to calculate the similarity between one data points to another.

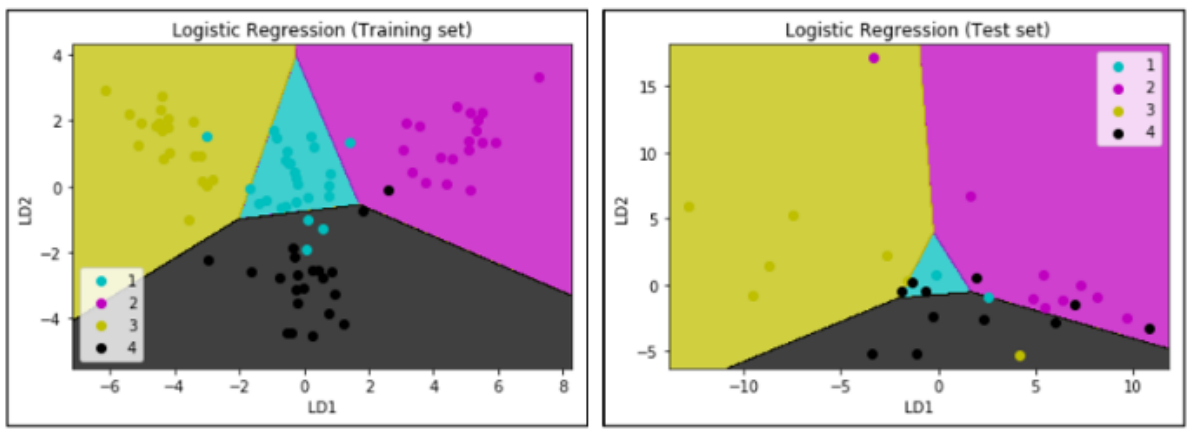

**Figure 5:** Dimensionality reduction through LDA applied on statistical features

## 3.7 Model Specification

Several supervised classification models have been applied to detect affective states. However, the random forest outperforms the XGboost, Naïve Bayes (NB), Gradient boosting classifier, and Multi-layer perceptron.

### 3.7.1 Random Forest

Random forest is a decision tree based classifier with large numbers of decision trees are used as the classifiers. Each and every trees are constructed from bootstrapped dataset. The majority voting is applied as the ensemble mechanism to decide the final class.

The GINI impurity (G) is used to determine best splitting criteria (**Equation 6**).

$$G = \sum_{i=1}^{C} P(i) * (1 - P(i))  \quad\quad\quad\quad\quad (6)$$

Here, C is the number of classes and P ( ) is the probability of picking a data point of class i.

By employing the method of Holdout, our data is divided into two sets: Training and Test/Validation set i.e. a holdout set. We then trained the model on the training dataset and evaluate the model on the Test/Validation dataset. Typically, the training dataset is bigger than the holdout dataset. Typical ratios used for splitting the data set include 60:40, 80:20 and so on. Here, test and train is divided into 30/70. In K-fold, validation technique has a single parameter called k that refers to the number of groups that a given data sample is to be split (**Figure 6**).

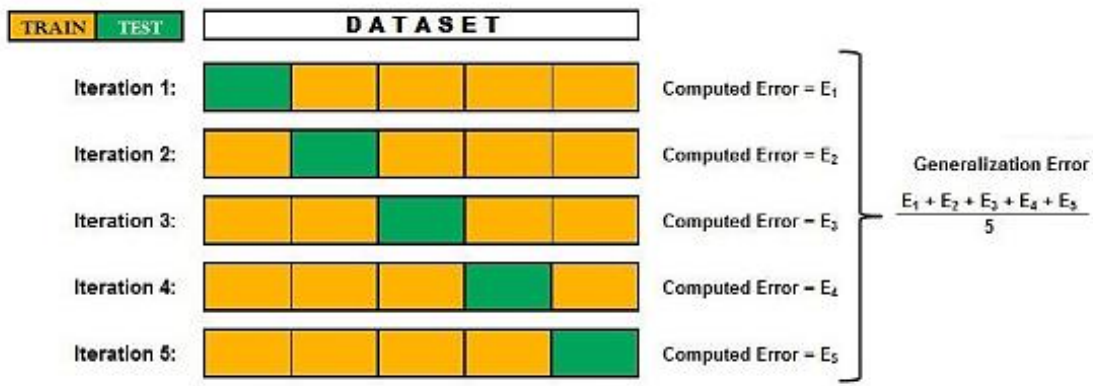

**Figure 6**: K-Fold Cross Validation Technique

However, in stratified K-fold Cross-validation, the limitation of K-fold validation techniques are minimized by introducing classes in our training and test dataset, for which we have used the Stratified k-Fold Cross-Validation.

## 4.0 Performance Evaluation

The overall study of emotion recognition and its features can be illustrated through the **Figure 1.** In statistical feature, we have applied XGB Classifier, Ridge Classifier CV, Ada Boost Classifier, Random Forest Classifier, SGD Classifier, Gaussian NB, Perceptron and others on unscaled statistical feature dataset. However, since none of them could produce accuracy to the expected level therefore we only mention the details about the best five (5) algorithms for our case **(Table 2)**.

**Table 2:** Training and testing accuracy of different classifiers applied on statistical features of EEG observations

| MLA Name | MLA Train Accuracy (%) | MLA Test Accuracy (%) |
|---|---|---|
| XGB Classifier | 100.00 | 86.21 |
| Ridge Classifier CV | 94.12 | 79.31 |
| Extra Tress Classifier | 100.00 | 75.86 |
| Linear SVC | 78.82 | 72.41 |
| Decision Tree Classifier | 100.00 | 65.51 |

In advanced feature **(Table 3)**, we have applied on unscaled advanced feature dataset. However, since none of them could produce accuracy to the expected level therefore we only mention the details about the best five (5) algorithms for our case. **Table 4** highlights the results of different standards of different classifiers applied on advanced features.

**Table 3:** Training and testing accuracy of different classifiers applied on advanced features of EEG observations

| MLA Name | MLA Train Accuracy (%) | MLA Test Accuracy (%) |
|---|---|---|
| AdaBoost Classifier | 52.56 | 85.19 |
| GaussianNB | 69.23 | 77.78 |
| Decision Tree Classifier | 100.00 | 74.07 |
| SGDC Classifier | 65.38 | 66.67 |
| XGB Classifier | 100.00 | 66.67 |

**Table 4:** Precision, recall and F1-score of different classifiers applied on advanced features.

| MLA Name | Precision (Macro) (%) | Precision (Micro) (%) | Precision (Weighted) (%) | Recall (Macro) (%) | Recall (Micro) (%) | Recall (Weighted) (%) | F1-score (Macro) (%) | F1-score (Micro) (%) | F1-score (Weighted) (%) |
|---|---|---|---|---|---|---|---|---|---|
| Extra Trees Classifier | 75.8929 | 75.8621 | 86.4532 | 81.4286 | 75.8621 | 75.8621 | 73.0769 | 75.8621 | 75.9861 |
| XGB Classifier | 68.75 | 72.4138 | 87.069 | 58.9286 | 72.4138 | 72.4138 | 62.8105 | 72.4138 | 78.3593 |
| Bagging Classifier | 70.00 | 72.4138 | 83.3333 | 67.8571 | 72.4138 | 72.4138 | 66.069 | 72.4138 | 76.2756 |

After combining both the statistical and advanced features, we have designed the **Table 5**.

**Table 5:** Training and testing accuracy of different classifiers applied on fused advanced and statistical feature

| MLA Name | MLA Train Accuracy (%) | MLA Test Accuracy (%) |
|---|---:|---:|
| Random Forest Classifier | 100.00 | 89.66 |
| XGB Classifier | 100.00 | 86.21 |
| Gradient Boosting Classifier | 100.00 | 82.76 |
| Perceptron | 59.77 | 79.31 |
| Gaussian NB | 71.26 | 79.31 |

As discussed earlier that we have applied XGBoost and mRMR on advanced and statistical feature dataset for data selection. The mentioned two techniques would select the features that have impact on emotion accuracy and neglect the excess features. We combined those selected features of advanced and statistical dataset and applied machine-learning algorithm on them. For all the cases, epoch is 10,000. Earlier we have applied the machine-learning algorithm on combined dataset of advanced and statistical features and found accuracy more than 86%. Now applied the machine learning algorithms on the dataset of selected features from both the dataset, and we achieved accuracy more than 86%. Therefore, it is quite evident that by using the selected features only, we can achieve the same accuracy, which means that by using lesser features in lesser time we may achieve the expected accuracy. In the same way, we have also applied the above mentioned classifiers and others on unscaled advanced and statistical combined selected feature dataset. Since, all of them could not produce accuracy to the expected level therefore we only mention the details about the finest Four (4) algorithms for our case. We have selected features from fusion of statistical and advanced feature using mRMR. From PCA, we observe that around 30 features are required to achieve 100% variance. However the case is little different in advanced feature dataset where the numbers of accurately identified cases have reduced to some extent due to lesser contribution of advanced feature to the system. However, the scenario will change when there is fusion in advanced and statistical features. A Receiver Operator Characteristic (ROC) curve is a graphical plot used to show the diagnostic ability of binary classifiers. The ROC curves in statistical dataset show that the curve are in the top-left corner which implies the better performance of the dataset. However the ROC curves in advanced feature dataset were not as satisfactory as of statistical features. There are fluctuations from gross accuracy and could not help to produce fine area under the curve. When the advanced and statistical features are fused together in the ROC curve, they produce performance better than those of advanced features **(Figure 7)**.

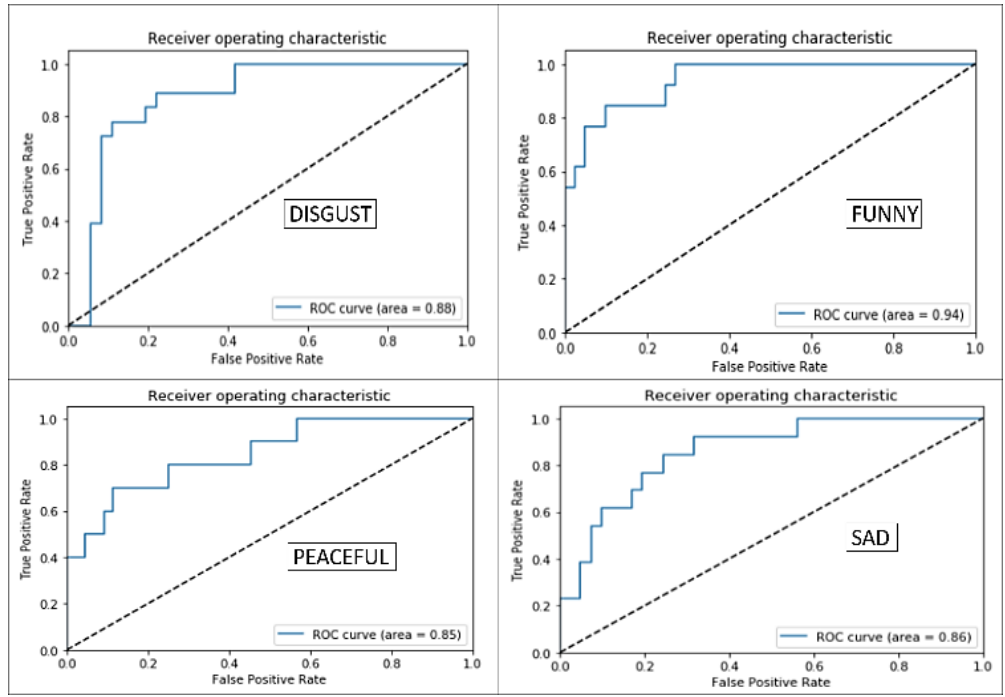

**Figure 7:** ROC curve of SAD, HAPPY, DISGUST and PEACEFULL affective states.

## 5.0 Conclusion

The versatility and complex human nature have made the human affective state recognition as a challenging problem in affective computing domain. However, this research used two channel EEG brain waves for real-time recognition of human affects. The EEG signal is collected from 100 participants where 15 video stimuli of different affects are used. The fusion of advanced and statistical features are used to train the affective state recognition system. The strategy of feature level fusion enhanced the accuracy level and finally reached to 89.66%. This accuracy level is more than the depth feature based emotion recognition approach of 87.5% accuracy. It is also better than radio frequency based emotion analyser (72%) and SVM based emotion classifier (82.9%). As the affect recognition is increasingly used in different kinds of games and virtual reality so the proposed affective state recognition system could be used to give players more natural control over their social avatars.